# Landau theory and self-assembly of spherical nanoclusters and nanoparticles with octahedral symmetry

D.V. Chalin[1] and S. B. Rochal[1]

[1] *Faculty of Physics, Southern Federal University, 344090 Rostov-on-Don, Russia*

Spherical nanoclusters and nanoparticles are rising materials whose functional design provides many useful applications ranging from catalysis, molecular sensing, gas storage to drug targeting and delivery. Here, we develop phenomenological crystallization theory of such spherical structures with octahedral symmetries $O$ and $O_h$. Within the developed theory, we propose a method, which is based on constructing irreducible octahedral density functions and allows to predict the positions of structural units in the spherical nanoobjects. The proposed theory explains the structures of the simplest known metal nanoclusters, some metal-organic polyhedra and membrane protein polyhedral nanoparticles, and also predicts more complex chiral spherical structures and achiral assemblies characterized by the geometry of semiregular polyhedra. A relationship between the constructed irreducible octahedral functions and spherical lattices, obtained by mapping a plane hexagonal order onto a spherical surface through an octahedron net, is discussed as well.

## I. INTRODUCTION

The principles of Landau phenomenological theory [1,2] were laid down almost nine decades ago. Since then, the Landau approach has remained a primary tool for the analysis and interpretation of experimental data on the behavior of physical systems in the vicinity of a phase transition (PT). Initially, this theory was proposed to explain second-order structural phase transitions of group-subgroup type in crystals [1,2]. Such transitions are usually associated with either small displacements of some atoms, or small changes in the probability of atoms to occupy certain positions. However, the universality of the mathematical apparatus, which includes the construction and analysis of an invariant potential or functional depending on the critical multicomponent order parameter (OP), allowed to describe various phase transitions in completely different physical systems. As a vivid example, one can recall that historically the first theory of superconductivity proposed by Landau and Ginzburg used the condensate wave function [3] as a physical implementation of the OP whose square modulus is proportional to the probability to detect a Cooper pair.

In the last half century, Landau theory has been repeatedly used to explain features of crystallization and structures of both well-known and newly discovered systems. For example, the authors of the well-known work [4], considering the 3rd degree invariant terms in the Landau functional, showed that during the crystallization of an isotropic melt, the structure with a body-centered cubic (BCC) lattice should be the most energetically favorable in the crystalline phase. An interesting version of the weak crystallization theory, describing a transition close to a second-order one, was proposed by Brazovskii [5]. Subsequently, the theory was further developed to explain the thermodynamic behavior and arrangement of various liquid crystal phases [6,7].

Landau theory allowed to explain the existence and stability of other interesting metallic systems – quasicrystals, discovered in 1982 [8]. Unlike crystals, these systems can possess 5-, 8-, 10- and 12-fold axes that are not compatible with crystallographic rotational symmetry, but at the same time, as in crystals, they have a long-range order. The first discovered example of a quasicrystal, which was the rapidly cooled AlMn alloy [8], had an icosahedral symmetry. The subsequent papers [9,10] proposed options for constructing the Landau functional, which favors energy-wise an icosahedral quasicrystalline structure. It is worth noting that additional Goldstone degrees of freedom in quasicrystals corresponding to gapless phason modes [11] can be understood in terms of the Landau approach as well [9].

Another example of interesting nanoassemblies, whose structures can be described within the framework of Landau theory, is the protein shells (capsids) of small viruses. These structures can have a spherical shape and very often adopt the icosahedral point symmetry $I$. Considering the crystallization of density waves with the same symmetry on a spherical surface [12,13], the authors proposed a simple and efficient method for calculating the mass centers of the proteins that form a viral shell and explained the existence of a series of abnormal viral capsids, contradicting the pioneering theory which provides the structural classification of capsids [14]. Fifteen years ago, when the work [12] was written, very little was known about the thermodynamics of self-assembly of viral capsids, and based on the absence of a cubic invariant in the Landau potential, the authors [12] suggested that the considered transition could be of the second order. Subsequently, the thermodynamic part of the work [12] was subjected to quite fair criticism [15], however, in our opinion, the predictions [12,13] of possible simplest spherical structures with icosahedral symmetry $I$ are quite correct and still relevant.

Various cubic, octahedral and more complex [16–18] nanoparticles with $O_h$ symmetry have been known for a long time. Over the past couple of decades, the development of synthesis techniques has allowed to design novel compact structures in which the positions of structural units (SUs) form regular and semiregular polyhedra with octahedral symmetry. These structures include nanoclusters containing octahedral, cubic and cuboctahedral Au or Ag cores [19–21] surrounded by molecular complexes, as well as various colloidal polyhedral nanoparticles [22,23] and metal-organic polyhedra [24–26].

Simultaneously, interesting protein nanoparticles, showing promise in drug delivery and targeting, antigen display, vaccination, and other fields, were synthesized [27]. In 2014, it was experimentally shown that 24 proteins on a spherical lipid membrane can form a chiral structure with the geometry of a snub cube (a well-known Archimedean solid with $O$ symmetry) [28]. A few years later, it was discovered that α- and γ-tocopherol transfer proteins (TTPs) [29,30], which regulate vitamin E transport in mammals, also self-assemble into stable chiral structures of 24 monomers that form a snub cube (see Fig. 1).

To date, a microscopic theory explaining the formation of such nanoclusters from 24 SUs has been proposed [31].

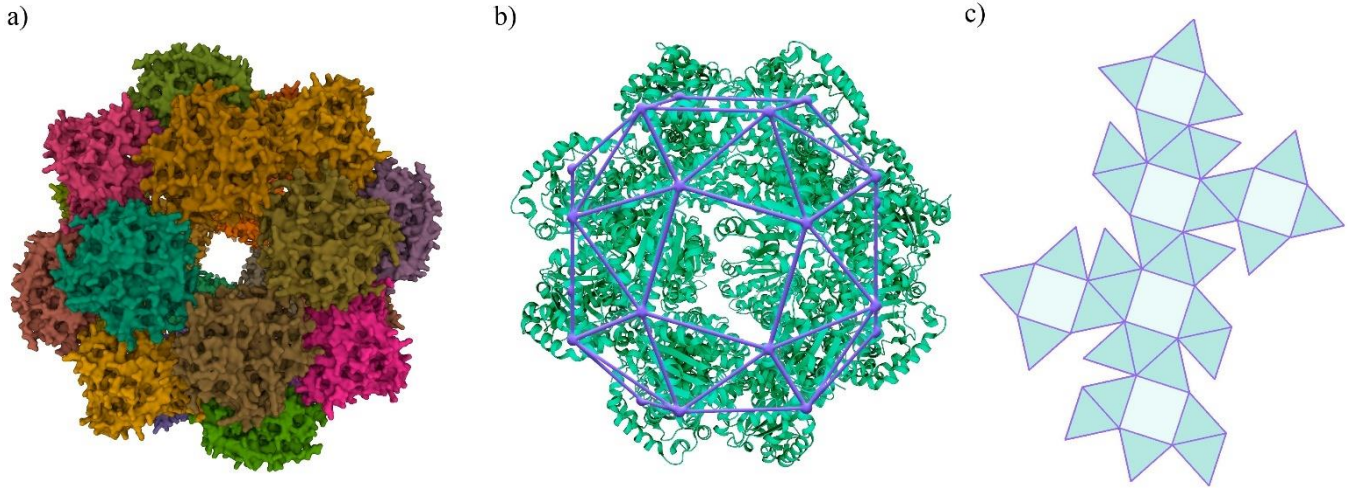

FIG. 1. (a) Model of a nanocluster consisting of 24 α-TTP monomers, see ID code 5MUG in Protein Data Bank [32]. (b) Cartoon representation of a given protein cluster superimposed with a snub cube. (c) The snub cube net.

In this paper, developing and revising the approach [12], we construct a general phenomenological theory of self-assembly and discuss the structures of the simplest chiral and achiral spherical structures with octahedral symmetries $O$ and $O_h$, respectively. We propose a simple method that predicts positions of SUs in spheical nanoparticles and establish a relationship between calculated structures and spherical lattices obtained by mapping a planar hexagonal order onto a sphere through octahedron nets.

## II. THEORY OF CRYSTALLIZATION ON A SPHERICAL SURFACE

To consider the thermodynamics of assembly of spherical nanostructures, we follow the main principles of Landau phenomenological theory. The positions of SUs (atoms or molecules) that form a nanocluster are described by the statistical density distribution function $\rho$. The thermodynamics of crystallization (as well as other structural phase transitions) is described using a non-equilibrium potential, which is an expansion of the free energy into a series in OP components in the vicinity of a phase transition. The coefficients of terms in this expansion depend on external parameters such as temperature and pressure, while the terms themselves are invariant in the space of the OP components with respect to the symmetry group $G_0$ of the initial high symmetry phase [33].

Within the framework of the classical Landau theory, the density distribution (in the vicinity of a structural phase transition) can be presented as
$$\rho = \rho_0 + \delta\rho, \qquad (1)$$
where $\rho_0$ is the microscopic density of the initial phase, $\delta\rho$ is the density variation that linearly depends on the OP components and describes the ordering of SUs. A phase transition induces spontaneous symmetry breaking i.e., in the ordered low-symmetry phase, the variation $\delta\rho \neq 0$. This critical density $\delta\rho$ determines the structure of the ordered phase and is expressed as a linear combination of functions that span an irreducible representation (IR) of the group $G_0$. The coefficients of these functions play the role of a multicomponent OP. Very often, in the vicinity of a phase transition, only one critical IR is nonzero, and the contribution of the remaining (noncritical) degrees of freedom to the crystallization process is negligibly small. Therefore, the functions that span an IR of the critical OP are called a critical system of density waves (CSDW).

The same principles can be applied to consider the crystallization of spherical nanostructures. We describe the initial symmetry of the disordered phase by the continuous isotropic group $O(3)$. Then, the structure crystallizing from an isotropic medium will have the same symmetry as $\delta\rho$. Accordingly, near the crystallization point, the positions of SUs in the new ordered phase can be considered as coinciding with the positions of the maxima or minima of this function.

Irreducible representations of $O(3)$ are constructed as products of IRs of the groups $SO(3)$ and $C_i$. The latter contains only the identity element $\hat{E}$ and the spatial inversion $\hat{I}$ and, consequently, has only two one-dimensional IRs: symmetric representation $A_g$ and antisymmetric one $A_u$. Functions spanning the latter IR change the sign under the inversion $\hat{I}$. Note that the same transformational properties are exhibited by pseudoscalars (for example, the scalar triple product of different vectors). However, if we take into account that $\delta\rho$ describing the density of the ordered phase is an ordinary scalar function of a radius vector, then it can only be expressed in terms of the spherical harmonics $Y_{lm}$, which span IRs of $O(3)$ with parity $(-1)^l$. Let us denote these IRs as $D_g^{(l)} \equiv D^{(l)} \otimes A_g$, where $D^{(l)}$ is an irreducible representation of $SO(3)$ also spanned by $Y_{lm}$.

In the present paper, we consider only those clusters whose symmetry lowers to an octahedral one ($O$ or $O_h$). Naturally, the crystallization of such structures can be driven by IRs with different $l$. For each representation $D_g^{(l)}$ indexed by wave number $l$, the critical variation $\delta\rho_l$ is constructed using a set of functions forming CSDW as

$$\delta\rho_l = \sum_{m=-l}^{m=l} A_{lm} Y_{lm}(\theta, \varphi), \quad (2)$$

where $A_{lm}$ is the coefficients of spherical harmonics $Y_{lm}$; $\theta$ and $\varphi$ are azimuthal and polar angles in the spherical coordinate system.

The thermodynamic potential is written in the form, standard for the Landau theory: $F = F_0 + F_2 + F_3 + F_4 + \cdots$. The terms of this expansion are constructed using invariant combinations of CSDW coefficients with the same wave number $l$ and are explicitly written as

$$\begin{aligned}
F_2 &= A(T,P) \sum_{m=-l}^{m=l} A_{l,m} A_{l,-m}, \\
F_3 &= B(T,P) \sum_{m_1,m_2,m_3} a_{m_1,m_2,m_3} A_{l,m_1} A_{l,m_2} A_{l,m_3} \\
&\quad \times \delta(m_1 + m_2 + m_3), \\
F_4 &= \sum_k C_k(T,P) \sum_{m_1,m_2,m_3,m_4} a^k_{m_1,m_2,m_3,m_4} \\
&\quad \times A_{l,m_1} A_{l,m_2} A_{l,m_3} A_{l,m_4} \delta(m_1 + m_2 + m_3 + m_4),
\end{aligned} \quad (3)$$

where $a_i$ is the weight coefficients of the group $O(3)$ (for example, Clebsch-Gordon coefficients for the third-order term $F_3$), $\delta(0) = 1$, $\delta(i \neq 0) = 0$, $A(T,P)$, $B(T,P)$, $C_k(T,P)$ are the Landau expansion coefficients, depending on the temperature $T$ and pressure $P$.

Since the considered critical density variation $\delta\rho_l$ has parity $(-1)^l$, crystallization of the chiral structures with the symmetry $O$ can be driven only by the representations with odd $l$, while crystallization of the structures with the symmetry $O_h$ can be driven only by IRs with even $l$. We also note that for any odd number $l$, the term $F_3$ and all other odd terms in the Landau expansion vanish identically, which essentially means that crystallization can be a second-order transition. However, when considering the potential of a higher degree (and the corresponding microscopic interactions), the considered transition may well be of the first order, as it occurs, for example, in the case of the transition from para- to ferroelectric phase in a significant number of ferroelectrics, for which the OP is the polarization vector, and there are no invariants of odd degree in the Landau expansion [34]. At the same time, for IRs of $O(3)$ with even $l$, one can always construct a cubic invariant, and crystallization should be a first-order transition for this reason alone [35].

The choice of the OP should be detailed even more since the function $\delta\rho_l$ with octahedral symmetry can be constructed not for all, but only for certain wave numbers $l$. For this purpose, we restrict the IRs of $O(3)$ to the considered octahedral groups and select only those ones satisfying the Birman criterion [36], then the allowed values of $l$ form the following sequence: $l = 4,6,8,9,10,12,13,14, \ldots$. The analysis based on the theory of invariants (see Appendix) shows that any critical OP driving the assembly of a spherical nanocluster must have a wave number $l$ that satisfies the following relationship

$$l = 9i + 4j + 6k, \quad (4)$$

where $i = 0,1$; $j$ and $k$ are positive integers. Solutions of Eq. 4 with $i = 1$ and $i = 0$ define IRs driving the crystallization of chiral and achiral structures, respectively. Also note that the number of different solutions $(j,k)$ of Eq. 4 for a given wave number $l$ is equal to parametricity of $\delta\rho_l$ in the low-symmetry phase. By the parametricity of the critical density $\delta\rho_l$ (potentially minimizing the Landau functional), we mean the number of linearly independent amplitudes $A_{lm}$ in Eq. 2. Thus, the parametricity is equal to the number of mutually orthogonal basis functions $f_l^\alpha(\theta,\varphi)$ spanning totally symmetric representation of the group $O(3)$, which drives the assembly of the nanostructure. If the parametricity $\eta$ is equal to 1, then $\delta\rho_l$ is defined up to an arbitrary multiplier and does not contain any internal variables. In the general case, $\delta\rho_l$ is expanded in terms of a set of totally symmetric functions as

$$\delta\rho_l(\theta,\varphi) = \sum_{\alpha=1}^{\eta} C_\alpha f_l^\alpha(\theta,\varphi), \quad (5)$$

The number $\eta$ can also be found by directly calculating the number of times the considered representation of $O(3)$ subduces the identity representation [37]:

$$\eta = \frac{1}{|G|} \sum_G \xi(\hat{g}), \quad (6)$$

where for both considered cases of groups $O$ and $O_h$, the summation can be carried out over only elements $\hat{g}$ of the group $O$, $|G|$ is the group order equal 24, $\xi(\hat{g})$ are the characters of the symmetry group $O(3)$ elements. They are found as

$$\xi(l,\alpha) = \sin\frac{[(l+1/2)\alpha]}{\sin(\alpha/2)},$$

where $\alpha$ is the rotation angle defined by the element $\hat{g}$. Direct calculation using Eq. 6 yields

$$\eta = \frac{1}{24}\left[2l + 1 + 6\xi\left(l,\frac{\pi}{2}\right) + 8\xi\left(l,\frac{2\pi}{3}\right) + 9\xi(l,\pi)\right]. \quad (7)$$

It is interesting to note that for even $l \leq 10$ and for odd $l \leq 19$, the expansion (5) contains only one function $f_l(\theta,\varphi)$. One can directly verify this result using Eqs. 4 and 7. Accordingly, the density of the ordered phase is found as $\delta\rho_l = Cf_l(\theta,\varphi)$, where $C$ is an arbitrary constant. Thus, the positions of SUs in crystallizing spherical nanoclusters are determined only by the positions of maxima/minima of the density wave $f_l(\theta,\varphi)$ and do not depend on the value of $C$. Let us call the functions $f_l(\theta,\varphi)$ as irreducible single-parameter octahedral density functions. Since the parity of $\delta\rho_l$ directly depends on the index $l$, the explicit form of $f_l(\theta,\varphi)$ (for both $O$ and $O_h$ symmetries) can be obtained by averaging the spherical harmonics $Y_{lm}(\theta,\varphi)$ over the group $G = O$:

$$f_l(\theta,\varphi) = \frac{1}{|G|} \sum_G \hat{g} Y_{lm}(\theta,\varphi). \quad (8)$$

For a fixed number $l$, the averaging procedure (8) for different values of $m \in [-l,l]$ gives either 0 or, up to a complex multiplier, the same function $f_l(\theta,\varphi)$. The result of the averaging procedure (8) can be always made real by choosing an appropriate multiplier. For functions $Y_{lm}$ with $m$ being multiple of 4, the averaging (8) never yields zero since these functions are invariant with respect to rotations about 4-fold axis. If Eq. (4) has more than one solution, then the components of the multicomponent density function can be found by averaging $Y_{lm}$ with different $m$. Concurrently, while finding non-zero averaged functions, one can use an algorithm for constructing the basis of an orthogonal irreducible subspace, the dimension of which will be equal to the number of solutions of Eq. (4). Leaving this rather cumbersome analysis for a future study, below, we consider only the simplest single-parameter CSDWs.

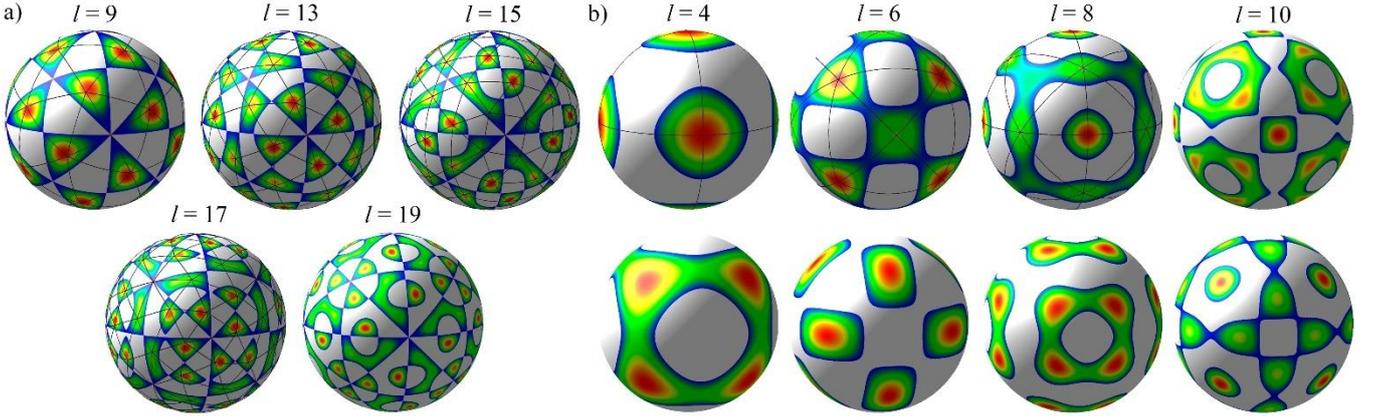

FIG. 2. Irreducible single-parameter density functions. Panel (a) shows density waves $f_l(\theta,\varphi)$ with wave numbers $l=9,13,15,17,19$ describing the crystallization of chiral spherical structures with symmetry $O$. Panel (b) shows density waves $f_l(\theta,\varphi)$ with numbers $l=4,6,8,10$ corresponding to achiral structures with symmetry $O_h$. The minima and maxima of the functions $f_l(\theta,\varphi)$ with even $l$ describe completely different self-assemblies, so the top row of the panel (b) shows only the part of the functions $f_l(\theta,\varphi)>0$, while the bottom row shows the part $f_l(\theta,\varphi)<0$. Black thin lines show primitive spherical lattices constructed using octahedron nets and superimposed on the density functions with $l=4,6,8,9,13,15,17$.

Figure 2 demonstrates all the functions describing the self-assembly of spherical structures with octahedral symmetry. The color change from blue to red corresponds to a growth of the function $f_l(\theta,\varphi)$. Panel (a) shows the antisymmetric functions $f_l(\theta,\varphi)$ with an odd wave number $l$: under the spatial inversion and the mirror planes of the octahedral group $O_h$, they change their sign. Accordingly, the positive part of the irreducible function ($f_l(\theta,\varphi)>0$) and its negative part ($f_l(\theta,\varphi)<0$) describe the crystallization of enantiomorphic "right" and "left" chiral objects. Nonequilibrium Landau potentials with odd $l$ are invariant under the transformation $\delta\rho_l \to -\delta\rho_l$, and the corresponding equilibrium states have the same energy and can be considered as domains. In panel (a) of Fig. 2, we show only the part $f_l(\theta,\varphi)>0$. Note that the positions of the maxima of the antisymmetric function $f_l(\theta,\varphi)$ have trivial symmetry and, therefore, their number is a multiple of 24, which is the number of elements in the group $O$. Symmetrically nonequivalent sets of maxima form regular orbits of the group $O$. The number of such orbits for CSDWs with the wave numbers $l=9,13,15,17,19$ is equal to $N=1,2,3,3,5$, respectively.

Figure 2b shows density waves $f_l(\theta,\varphi)$ with the even wave numbers $l=4,6,8,10$. In this case, the nonequilibrium Landau potentials contain invariants of odd degrees (including the cubic ones), and the critical variations $\delta\rho_l$ corresponding to the functions $f_l(\theta,\varphi)$ and $-f_l(\theta,\varphi)$ describe the assembly of different equilibrium phases with the same symmetry. In Figure 2b, the top row shows the positive part of the irreducible octahedral functions, which we denote as $f_l^+(\theta,\varphi)$, and the bottom row shows the negative part denoted as $f_l^-(\theta,\varphi)$. To draw $f_l^-$, we inverted the sign of functions $f_l$. Since the even harmonics $f_l(\theta,\varphi)$ are invariant with respect to the cube mirror planes and the spatial inversion, maxima and minima of these functions can be located in highly symmetrical positions (at the 2-,3- of 4-fold axes or the cube mirror planes).

Interestingly, the critical density waves in Figure 2 with small $l$ describe the structures of regular and semiregular polyhedra. For example, the positions of maxima of the function $f_9$ form the vertices of an Archimedean solid called the snub cube. The density waves $f_l^-$ with $l=4,6,8$ in panel (b) correspond to a cube, a cuboctahedron, and a rhombicuboctahedron, while the density wave $f_4^+$ describes an octahedron. The irreducible functions $f_6^+$, $f_8^+$, $f_{10}^+$, and $f_{10}^-$ have two sets of symmetrically nonequivalent maxima i.e., two regular orbits of the group $O_h$. These orbits also correspond to the vertices of Platonic and Archimedean solids such as cuboctahedron, rhombicuboctahedron, and truncated octahedron (see density waves $f_8^+$, $f_{10}^+$, and $f_{10}^-$ in Figure 2b, respectively).

Finally, let us note that the functions $f_4^+$, $f_9$ and $f_{13}$ correspond to the solutions of the Tammes problem [39,40], which are, respectively, the densest packings of 6, 24 and 48 identical disks of maximum radius on the sphere. The same functions also correspond to the solutions of the Thomson problem [41,42] that considers the most energetically favorable configurations of charged particles retained on the sphere.

## III. DISCUSSION AND CONCLUSION

In the present section, we describe the relationship between the calculated irreducible density functions and real structures with octahedral symmetry and consider problems related to the thermodynamics of self-assembly of the studied objects. However, first, let us discuss a relationship between the theories describing the crystallization on planar and spherical surfaces. Like a plane, a sphere is also a two-dimensional manifold, and, therefore, between planar lattices and spherical structures, as objects obtained within the same Landau theory, there must be a relationship, which we establish below.

It is well known that the Landau theory in its simplest form, when considering crystallization on a planar surface, gives a solution, which is a superposition of three plane waves with wave vectors $q_i$ directed from the center to the vertices of a regular triangle [4]. The corresponding nonequilibrium potential has a cubic term and the following solution:

$$\delta\rho_l(r) = C \sum_{i=1}^{3} \cos(q_i r), \quad (9)$$

where $r$ is a two-dimensional radius vector. The maxima and minima of the critical density (9) correspond to trigonal and honeycomb lattices, depending on the sign of the coefficient $C$.

Any hexagonal periodic order can be projected onto a sphere using a net of a regular polyhedron with triangular faces: the polyhedron net is mapped onto the hexagonal order in such a way that its vertices are superimposed with the six-fold axes, while its edges become the translations of the hexagonal order (see Fig. 3). This mapping procedure ensures smooth "gluing" of the net everywhere, except for the axes of the polyhedron, where topological defects (so-called disclinations [38]) are formed.

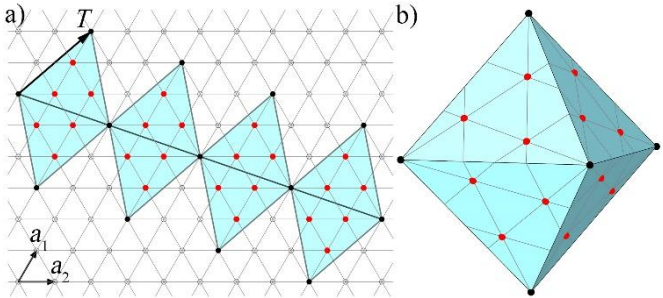

FIG. 3. Smooth mapping of the primitive hexagonal order onto the octahedron surface using the octahedron net. The case when the octahedron edge coincides with the translation $T = (2,1)$ is illustrated. After the mapping, the nodes located on the octahedron faces (shown in red) have trivial symmetry, while the nodes located at the octahedron vertices have $C_4$ symmetry.

Naturally, a six-fold axis cannot be preserved at the polyhedron vertex, but all other nodes of the resulting spherical lattice still preserve local hexagonal symmetry. Similar procedures can be performed using nets of regular icosahedron, octahedron, and tetrahedron; the mapping itself can be indexed by two integers $(h, k)$, which define a hexagonal order translation $T$ transformed into an edge of the polyhedron net:

$$T = ha_1 + ka_2 \equiv (h, k),$$

where $a_1$ and $a_2$ are the basis translations of the hexagonal Bravais lattice.

Spherical lattices are obtained by radially projecting polyhedra onto a sphere. Polyhedra with vertices coinciding with the nodes of such spherical lattices are known as geodesic polyhedra. As a striking example of structures based on geodesic icosahedra, one can recall capsids of some icosahedral viruses [14,43]. In the case of a honeycomb lattice mapped onto the surface of a spherical octahedron, the resulting structures consisting of hexagons and tetragons are dual to geodesic octahedra: their nodes, just as in the planar case, coincide with the centers of the triangles that form the primitive spherical lattices. Note that many fullerene molecules have the structure of analogous spherical polyhedra formed by hexa- and pentagons [44].

Below we consider spherical structures with octahedral symmetry. It is easy to see that the locations of the maxima of the functions $f_4^+, f_6^+$ and $f_8^+$ (see Fig. 2b) correspond to primitive (triangular) spherical lattices $(1,0)$, $(1,1)$, and $(2,0)$, respectively. Note that, in a such-type spherical lattice, every node, except for the ones located in octahedron vertices and surrounded by 4 nodes, has 6 neighboring nodes.

To rationalize the arrangements of structures generated by the maxima of the functions $f_4^-$, $f_6^-$ and $f_8^-$, we, using octahedron nets, projected these structures onto the plane. Fig. 4 shows possible correspondences between the constructed nets and a planar honeycomb lattice. The nets shown are characterized by the same indices $(h, k)$ as the above discussed primitive spherical lattices. However, in the cases (b) and (c), the honeycomb rings are defective and the relation between the considered $f^-$ and $f^+$ spherical functions is more complex than the well-known duality between the honeycomb and primitive planar lattices.

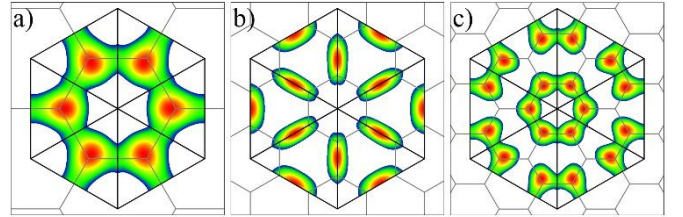

FIG. 4. Octahedral spherical functions $f_4^-$, $f_6^-$, and $f_8^-$ (panels (a), (b) and (c), respectively) projected on the plane. Symmetrically equivalent triangles with bold black edges correspond to the faces of an octahedron. Two triangles are added, and the 4-fold axis of octahedron is converted into the 6-fold one. The honeycomb lattice is superimposed in such a way that the octahedron edges are simultaneously the translations of this lattice.

In the case of chiral structures, the relation between the antisymmetric functions $f_9$, $f_{13}$, $f_{15}$, $f_{17}$ and the spherical lattices $(2,1), (3,1), (3,2), (4,1)$ is a bit more intricate: the nodes of the primitive triangular lattice, coinciding with the symmetry axes of an octahedron, now correspond to the saddles of spherical functions, and these functions vanish identically at these points. Therefore, in the chiral lattices superimposed on the density functions in Fig. 2a, the nodes lying on the symmetry axes of the octahedron do not correspond to the maxima of density functions, and asymmetric SU cannot be placed there. The remaining nodes correspond well to the positions of maxima of the functions $f_9$, $f_{13}$, $f_{15}$, and $f_{17}$. Note that distances between the nearest nodes with trivial symmetry are approximately equalized by the algorithm used in Ref. [43] for icosahedral spherical lattices when modelling viral capsids. Before the alignment, the correspondence between the maxima and the nodes of spherical lattices is slightly worse. It is interesting to note that the maxima (or minima) of icosahedral CSDWs with the first permitted odd $l$ values (see Ref. 12) coincide with the nodes of icosahedral spherical lattices [43], as in the considered case of octahedral CSDWs with $l = 9, 13, 15, 17$ (see Fig. 2). The solution with $l = 19$ is not a spherical lattice, while the solutions with greater $l$ become multicomponent. A similar breaking in the correspondence between primitive spherical lattices and CSDWs with increasing $l$ occurs in the icosahedral case as well.

Now, let us discuss the connection between real structures and the obtained single-parameter CSDWs. To date, a great variety of spherical nanoparticles and nanoclusters have been synthesized. The increased attention to these objects in material science can be explained by the vastness of their potential applications. On the one hand, colloidal nanoparticles with structures of the simplest regular polyhedra can find application in optoelectronics because of well-controlled plasmonic resonances [23]. On the other hand, metal-organic polyhedra can be used as nanocontainers with high selectivity, gas storage, sensors, and nanoscale reaction vessels [25].

The octahedral density wave $f_4^+$ with the smallest wave number, for example, describes the structure of well-known octahedral nanoclusters formed by Au atoms [45], as well as the structures of recently synthesized plasmonic octahedral nanoparticles [23]. The irreducible octahedral functions $f_4^+$, $f_4^-$ and $f_6^-$ describe well the SU positions in recently obtained colloidal nanoclusters [22], in which polysterene nanoparticles packed on organometallic core form an octahedron, a cube, and a cuboctahedron, respectively. The same functions can be used to explain the arrangements of SUs in many known octahedral, cubic, and cuboctahedral metal-organic polyhedra [24,26].

Speaking of chiral spherical structures, already in Ref. [12] devoted to self-assembly of spherical icosahedral capsids from individual asymmetric proteins, the authors has proposed an absolutely correct idea that asymmetric proteins can occupy only general positions with trivial symmetry, and if they are placed at the maxima (or minima) of an CSDW, then only the functions with odd $l$ are suitable to describe such an arrangement. Following the same logic, we can explain three experimentally observed protein structures [28-30] characterized by the geometry of a snub cube. The positions of SUs in these protein clusters are well described by the function $f_9$.

However, when considering the energies of the assembled spherical structures, this logic leads to a contradiction, similar to the one appearing in the theory [12] and noted in Ref. [15]. This contradiction manifests in the case of odd $l$, provided that the solution or melt consists of asymmetric molecules of the same chirality. Indeed, for an odd $l$, the simplest nonequilibrium potential is invariant with respect to the transformation $\delta\rho_l \to -\delta\rho_l$. However, if asymmetric proteins of only one chirality are present in the solution (which is usual for protein solutions), then only a structure of the corresponding chirality can be assembled from such proteins, and this structure, within the framework of the phenomenological theory, should obviously be more energetically favorable.

Criticizing Ref. [12], the authors of Ref. [15] proposed a theory based on the approach [5], in which the polynomial nonequilibrium Landau potential was replaced by a nonequilibrium functional with gradient terms. The functional from the very beginning was not invariant with respect to the transformation $\delta\rho_l \to -\delta\rho_l$. When searching for the corresponding solutions, it turned out that the expansion of the function $\delta\rho$ contains both even and odd spherical harmonics simultaneously. The expansion was further truncated by considering only the OP with the nearest symmetry-allowed wave numbers $l$. In this case, the corresponding Landau potential turned out to be reducible and included additional cube-linear terms that violated the 'extra' invariance.

We believe, such a correction to the original theory is quite reasonable; however, the use of a phenomenological functional, which includes differential operators for a system, in which the size of a protein is comparable to the size of the resulting nanocluster, is, in our opinion, an unreasonable overcomplication. Nevertheless, we absolutely agree that a more complex version of the theory may include several OPs with close allowed values of $l$. One can verify this with the following simple considerations. Even if we place point particles at the maxima/minima of any CSDW considered here and expand the resulting density function in spherical harmonics, then in such an expansion, functions with the nearest (symmetry-allowed) values of $l$ will also make a significant contribution up to tens of percent. And if the maxima/minima of an CSDW with an odd $l$ are occupied, then the expansion will also include the nearest even harmonics. Accordingly, the addition of new "mixed" invariants with appropriately chosen coefficients, can make an experimentally observed structure more energetically favorable. Thus, the transition from density functions to structures in which SUs occupy certain positions (CSDW maxima/minima), strictly speaking, corresponds to the utilization of more complex potentials with reducible OP.

In the considered octagonal case, the analysis within the group theory shows the existence of invariants, which are cubic in the functions $Y_{9,m}$ and linear in the functions $Y_{10,m}$ and $Y_{8,m}$. These invariants do not vanish in the chiral octahedral phase since the superposition of CSDWs $f_9$, $f_8$, and $f_{10}$ preserves the chiral symmetry $O$ of the function $f_9$. If the contribution of the functions $f_8$ and $f_{10}$ remains within 15%, then it has little effect on the positions of the maxima/minima of the function $f_9$, however, the presence of even density components breaks the 'extra' invariance of the nonequilibrium Landau potential (see Fig. 5). A similar consideration can be carried out for other chiral functions $f_l$

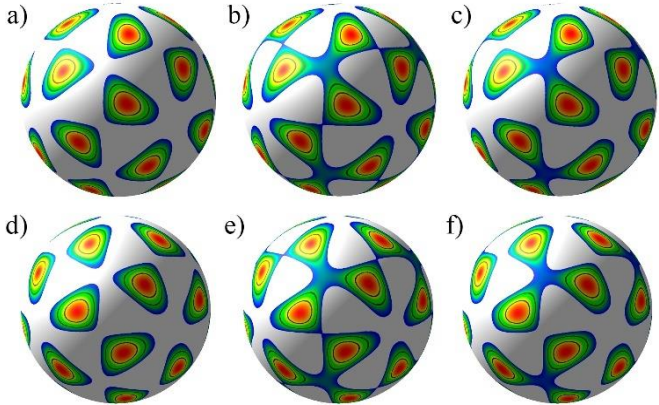

FIG. 5. Superpositions of single-parameter CSDWs $f_9$, $f_8$, and $f_{10}$ with different weights. The functions are normalized so that the integral of their square over the sphere surface is equal to 1. Panels (a)–(c) show mixed functions $f_9 + 0.15 f_8$, $f_9 + 0.1 f_{10}$ and $f_9 + 0.15 f_8 + 0.1 f_{10}$, respectively. For similar functions in panels (d)–(f), the contribution of the component $f_9$ is inverted: $f_9 \rightarrow -f_9$. Such a replacement of amplitudes leads to a change in the sign of the cube-linear invariant, but the mixed density functions, shown one above the other in the first and second rows, remain perfectly enantiomorphic.

The considered reducible OP is obviously not small since the interactions between its different irreducible components also play an important role in determining the energies of phases possible in the system. Thus, the construction of correct thermodynamics within the Landau theory requires consideration of potentials of a higher degree and with a large number of OP components, which makes this problem extremely complex, and results obtained within low degree potentials unreliable. For example, the fact that a chiral icosahedral solution with $l = 15$ is unstable in the case of the 4[th] degree Landau potential [15, 46] does not prove that this phase is, in fact, absent in models considering potentials of a higher degree.

Nevertheless, the solution of the above very complex mathematical task, which is necessary for constructing correct thermodynamics, may have no effect on the structural predictions presented in this paper. Let us emphasize that these predictions are based on the Birman criterion, which allows or forbids the realization of a phase for a given OP on the basis of purely symmetrical principles. As a result, the addition of new invariants to the Landau potential may not be essential for the arrangement of the resulting structures. Note that in this context, the structures of icosahedral viral capsids based on the icosahedron nets and corresponding to the spherical lattices (2,1), (3,1), (3,2) and (4,1) (in which the positions coinciding with the icosahedron symmetry axes are excluded) are actually observed in nature [43], despite the fact that a rigorous thermodynamic model based on the Landau theory and considering viral shells corresponding to the last three lattices has never been discussed. Moreover, there are many well-known icosahedral protein shells, arranged similarly to each of the spherical lattices mentioned above [43], and all the similar shells differ from each other only in slight variations in the locations of protein centers of mass.

In conclusion, the proposed crystallization theory for spherical structures with octahedral symmetry has allowed us to explain the structural organization of many known nanoparticles and nanoclusters. As an example of such structures one can recall well-known octahedral metallic nanoclusters [45], as well as more complex objects such as metal-organic polyhedra [24,26] and plasmonic nanoparticles [22,23], in which the positions of SUs form Platonic and Archimedean solids, namely, octahedron, cube, and cuboctahedron. Speaking of chiral objects, as far as we know, only 3 different protein nanoparticles [28–30] with a snub cube geometry (corresponding to the spherical lattice (2,1)) have been discovered, and we hope that our work will contribute to the search for new structures whose organization can be described by the other spherical lattices discussed. The theory developed in this paper for constructing irreducible octahedral functions and octahedral functions of general form on a sphere may also be of interest for other applications. Previously, similar icosahedral functions were used to study the charge distribution in proteins of viral capsids [47,48] and to analyse their shape [49,50].

## APPENDIX. THE INVARIANTS OF THE GROUPS $O$ AND $O_h$ IN 3D SPACE AND ON THE SPHERICAL SURFACE.

The following analysis within the framework of the invariant theory allows to establish the selection rule (4), which limits the possible irreducible representations driving the crystallization of spherical structures with octahedral symmetries $O$ and $O_h$. Let us consider the general properties of scalar functions, which are invariant with respect to the octahedral groups.

Any scalar function that is invariant under a point symmetry group can be formally represented as a polynomial series constructed from $x$, $y$, and $z$ components of the radius vector. For the group $O$, the terms of this series can be expressed in products of the following basis functions

$$J_0 = x^2 + y^2 + z^2,$$
$$J_1 = \prod_{i=1}^{4} \boldsymbol{n}_i^{(1)} \boldsymbol{r}, \quad J_2 = \prod_{i=1}^{3} \left(\boldsymbol{n}_i^{(2)} \boldsymbol{r}\right)^2, \quad J_3 = \prod_{i=1}^{9} \boldsymbol{n}_i^{(3)} \boldsymbol{r}, \quad (10)$$

where $\boldsymbol{r} = (x, y, z)$ is the radius vector, $\boldsymbol{n}_i^{(1)}$, $\boldsymbol{n}_i^{(2)}$ and $\boldsymbol{n}_i^{(3)}$ are the vectors along the 4 three-fold axes, 3 four-fold axes, and 9 two-fold axes of a cube, respectively

Let us stress that the properties of scalar functions that are invariant with respect to the group $O$ differ significantly from the properties of functions that are invariant under the group $O_h$, which includes the cube mirror planes and the spatial inversion. As is known, for groups generated by reflections, the number of basis invariants is equal to the dimension of the vector representation [51]. Therefore, the invariant basis, also known as integrity basis, generating the ring of invariant polynomials of the group $O_h$, includes only three functions: $J_0$, $J_1$ and $J_2$. In the case of the group $O$, the integrity basis also includes the function $J_3$, and, since the number of the basis functions exceeds the dimension of the vector representation, there is a functional dependence between the invariants (10). They form a syzygy: an algebraic equation of the 18th degree. Indeed, the square of $J_3$ can be represented as a polynomial function depending on $J_0$, $J_1$ and $J_2$. Therefore, any scalar polynomial function invariant with respect to the group $O$ cannot contain $J_3^2$ and must be linear in $J_3$. Thus, taking into account that the invariant $J_0$ is a constant on a sphere of fixed radius, Eq. 4 immediately follows from the properties of the integrity basis considered above.

## ACKNOWLEDGEMENTS

D.Ch. and S.R. acknowledge financial support from the Russian Science Foundation, grant no. 22-12-00105.